\documentclass[twocolumn,prl,reprint,superscriptaddress]{revtex4}
\usepackage{amsmath,amssymb}
\usepackage{graphicx}
\usepackage{float}
\usepackage{subfigure}
\usepackage{verbatim}

\usepackage{amsfonts}
\usepackage{dcolumn}
\usepackage{bm}
\usepackage{array,color}
\usepackage[colorlinks,citecolor=blue]{hyperref}
\usepackage[english]{babel}

\begin{document}

\title{Experimental quantification of asymmetric Einstein-Podolsky-Rosen steering}

\renewcommand{\thefootnote}{\fnsymbol{footnote}}

\author{Kai Sun$^*$}
\affiliation{Key Laboratory of Quantum Information, University of Science and Technology of China, CAS, Hefei, 230026, People's Republic of China}
\affiliation{Synergetic Innovation Center of Quantum Information and Quantum Physics, University of Science and Technology of China, Hefei, Anhui 230026, P. R. China}

\author{Xiang-Jun Ye\footnote{These two authors contributed equally to this work.}}
\affiliation{Key Laboratory of Quantum Information, University of Science and Technology of China, CAS, Hefei, 230026, People's Republic of China}
\affiliation{Synergetic Innovation Center of Quantum Information and Quantum Physics, University of Science and Technology of China, Hefei, Anhui 230026, P. R. China}

\author{Jin-Shi Xu}
\email{jsxu@ustc.edu.cn}
\affiliation{Key Laboratory of Quantum Information, University of Science and Technology of China, CAS, Hefei, 230026, People's Republic of China}
\affiliation{Synergetic Innovation Center of Quantum Information and Quantum Physics, University of Science and Technology of China, Hefei, Anhui 230026, P. R. China}

\author{Xiao-Ye Xu}
\affiliation{Key Laboratory of Quantum Information, University of Science and Technology of China, CAS, Hefei, 230026, People's Republic of China}
\affiliation{Synergetic Innovation Center of Quantum Information and Quantum Physics, University of Science and Technology of China, Hefei, Anhui 230026, P. R. China}

\author{Jian-Shun Tang}
\affiliation{Key Laboratory of Quantum Information, University of Science and Technology of China, CAS, Hefei, 230026, People's Republic of China}
\affiliation{Synergetic Innovation Center of Quantum Information and Quantum Physics, University of Science and Technology of China, Hefei, Anhui 230026, P. R. China}

\author{Yu-Chun Wu}
\affiliation{Key Laboratory of Quantum Information, University of Science and Technology of China, CAS, Hefei, 230026, People's Republic of China}
\affiliation{Synergetic Innovation Center of Quantum Information and Quantum Physics, University of Science and Technology of China, Hefei, Anhui 230026, P. R. China}

\author{Jing-Ling Chen}
\email{chenjl@nankai.edu.cn}
\affiliation{Theoretical Physics Division, Chern Institute of Mathematics, Nankai University, Tianjin, 30071, People's Republic of China}
\affiliation{Centre for Quantum Technologies, National University of Singapore, 3 Science Drive 2, Singapore, 117543}

\author{Chuan-Feng~Li}
\email{cfli@ustc.edu.cn}
\affiliation{Key Laboratory of Quantum Information, University of Science and Technology of China, CAS, Hefei, 230026, People's Republic of China}
\affiliation{Synergetic Innovation Center of Quantum Information and Quantum Physics, University of Science and Technology of China, Hefei, Anhui 230026, P. R. China}

\author{Guang-Can Guo}
\affiliation{Key Laboratory of Quantum Information, University of Science and Technology of China, CAS, Hefei, 230026, People's Republic of China}
\affiliation{Synergetic Innovation Center of Quantum Information and Quantum Physics, University of Science and Technology of China, Hefei, Anhui 230026, P. R. China}

\date{\today }

\begin{abstract}
Einstein-Podolsky-Rosen (EPR) steering describes the ability of one observer to nonlocally ``steer" the other observer's state through local measurements. EPR steering exhibits a unique asymmetric property, i.e., the steerability can differ between observers, which can lead to one-way EPR steering in which only one observer obtains steerability in the steering process. This property is inherently different from the symmetric concepts of entanglement and Bell nonlocality, and it has attracted increasing interest. Here, we experimentally demonstrate asymmetric EPR steering for a class of two-qubit states in the case of two measurement settings. We propose a practical method to quantify the steerability. We then provide a necessary and sufficient condition for EPR steering and clearly demonstrate one-way EPR steering. Our work provides new insight into the fundamental asymmetry of quantum nonlocality and has potential applications in asymmetric quantum information processing.
\end{abstract}

\maketitle

Quantum nonlocality, which does not have a counterpart in classical physics, is the characteristic feature of quantum mechanics. First noted in the famous paper published by Einstein, Podolsky and Rosen (EPR) in 1935 \cite{epr1935}, which aimed to argue the completeness of quantum mechanics, the content of quantum nonlocality has been greatly extended. In 2007, Wiseman \emph{et al.} summarized the different conditions of quantum nonlocality and reformulated the concept of steering \cite{wjd2007} originally introduced by Schr\"{o}dinger \cite{schr1935} in response to the EPR paper (usually referred to as EPR steering), which stands between entanglement \cite{epr1935} and Bell nonlocality \cite{bell1964} in the hierarchy. In the view of a quantum information task, EPR steering can be regarded as the distribution of entanglement from an untrusted party, whereas entangled states need both parties to trust each other, and Bell nonlocality is presented on the premise that they distrust each other \cite{jones2007,brunner2014}. As a result, some entangled states cannot be employed to realize steering, and some steerable states do not violate Bell-like inequalities. EPR steering provides a novel insight into quantum nonlocality, and it exhibits an inherent asymmetric feature that differs from both entanglement and Bell nonlocality. Consider two observers, Alice and Bob, who share entangled states. There are cases in which the ability of Alice to steer Bob's state is not equal to the ability of Bob to steer Alice's state. There are also situations in which Alice can steer Bob's state but Bob cannot steer Alice's state, or vice versa; these situations are referred to as one-way steering \cite{wjd2007,joseph2014}. Several theoretical \cite{criteria2009,midgley2010,walborn2011,wiseman2012pra,nava2012,chenjl2013,he2013,howell2013,brunner2014prl,paul2014,joseph2014,reid2014,reid2015,kogias2015,piani2015,kogias20152} and experimental studies \cite{saunders2010,wittmann2012,vitus2012,bennet2012,sun2014,seiji2015,sacha2015,cavalcanti2015,pan2015} have focused on the verification and applications of EPR steering. Experimental demonstrations of one-way steering with the measurements restricted to Gaussian measurements have been reported \cite{vitus2012,seiji2015}. A class of entangled qubit states that can be used to show the property of one-way steering with general projective measurements has been theoretically constructed \cite{joseph2014}, with the requirement that the weight of the entangled part of the states should be between 0.4983 and 0.5. To prepare this type of mixed entangled states, the biggest error bar of the weight should be less than 0.00085 which implies that the experimental requirement is high. There has been no experimental demonstration of this phenomenon until now.

In this work, we consider an EPR steering game between two observers, Alice and Bob, restricted to two-setting projective measurements. A value called steering radius $R$ is defined on the basis of steering robustness to quantify the steerability \cite{piani2015}. Asymmetric EPR steering and one-way steering are then clearly demonstrated for a class of two-qubit entangled states. For the case of non-steerability, we experimentally construct the local hidden state model (LHSM) \cite{wjd2007,joseph2014} and use local hidden states to reproduce the experimental results obtained in the steering process with high fidelity.

\begin{figure}
\centering
\includegraphics[width=0.47\textwidth]{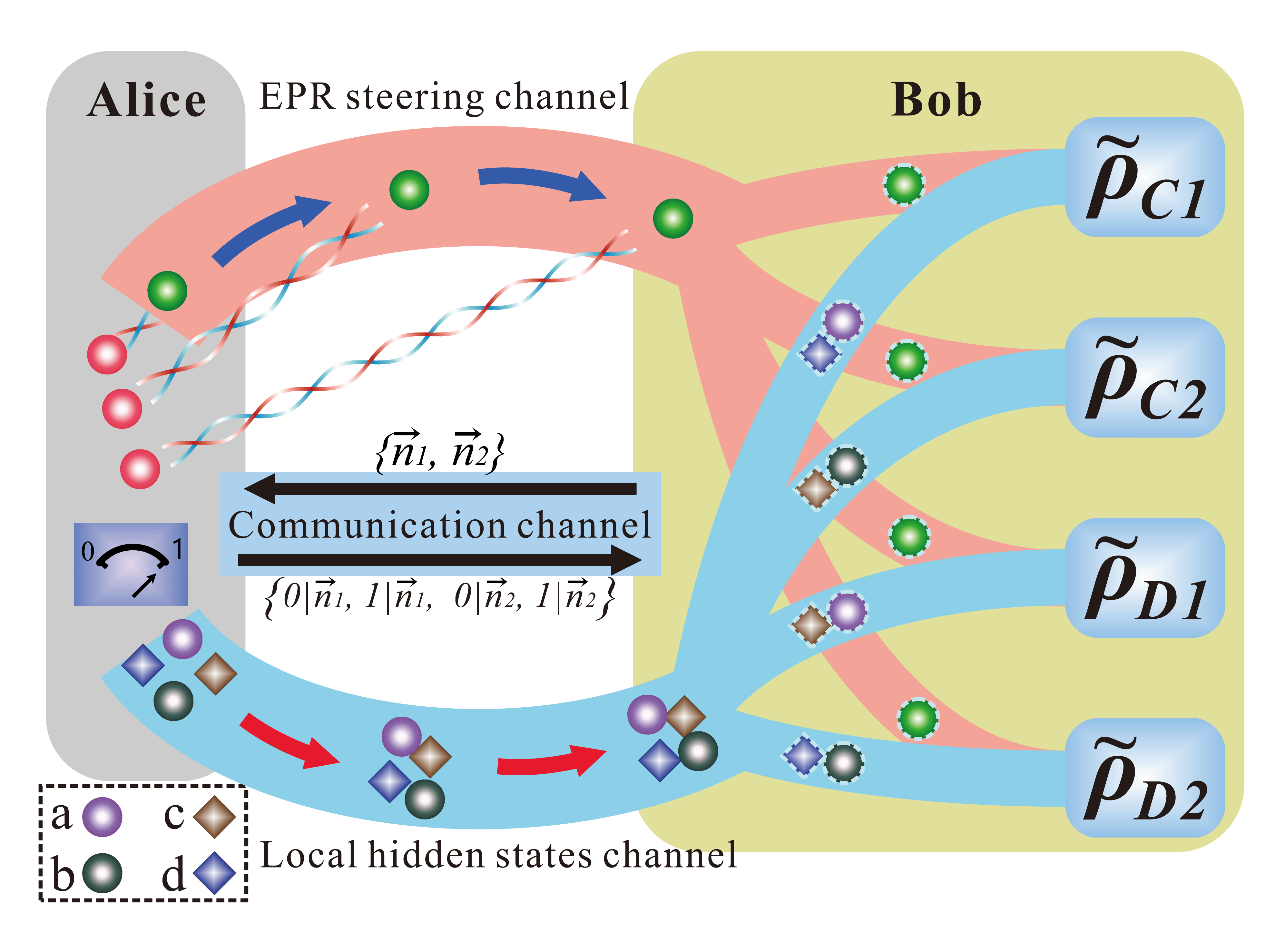}
\caption{The two-setting protocol of EPR steering and the strategy for the LHSM to reproduce the CSs in a failed steering task. Alice sends one of the two qubits to Bob through the EPR steering channel and measures her state along one of the two projective measurement settings $\vec{n}_1$ and $\vec{n}_2$ according to Bob's requirement. Alice then sends her measurement result $1|\vec{n}_1$ or $0|\vec{n}_1$ ($1|\vec{n}_2$ or $0|\vec{n}_2$) to Bob. Correspondingly, Bob obtains one of the four CSs, which are denoted as $\tilde{\rho}_{C1},\, \tilde{\rho}_{C2}$ (the two CSs for $\vec{n}_1$) and $\tilde{\rho}_{D1},\,\tilde{\rho}_{D2}$ (the two CSs for $\vec{n}_2$) after measuring his qubit (marked with a dashed frame). The classical communications between Alice and Bob occur in the communication channel. For a failed steering task, there exists an LHSM consisting of four local hidden states with the corresponding probabilities to reproduce the four CSs. The symbol labeled by $a$ ($b$, $c$, $d$) represents the local hidden state $\rho_a$ ($\rho_b$, $\rho_c$, $\rho_d$) with the corresponding probability $p_a$ ($p_b$, $p_c$, $p_d$). They are sent through the local hidden states channel. The combinations of two corresponding hidden states, represented by the dashed frame, are used to reconstruct Bob's four CSs.
}\label{theory}
\end{figure}

The steering process is illustrated via the EPR steering channel as shown in Fig. \ref{theory}. Alice sends one of the two particles to Bob and wants to persuade Bob to believe that she can steer his state. The analysis is the same when Bob wants to steer Alice's state. Bob's conditional states (CSs) obtained after receiving all results $\kappa|\vec{n}$ from Alice, where $\kappa|\vec{n}$ denotes that Alice gets the result $\kappa$ ($0$ or $1$) when measuring along the direction $\vec{n}$, can now represent as $\tilde{\rho}_{\kappa|\vec{n}}=\textup{Tr}_A[\rho_{AB}(\Pi_{\kappa|\vec{n}}\otimes I)]$ (unnormalized, where the normalized form is $\rho_{\kappa|\vec{n}}=\tilde{\rho}_{\kappa|\vec{n}}/\textup{Tr}[\tilde{\rho}_{\kappa|\vec{n}}]$) , where $\Pi_{\kappa|\vec{n}}=[I+(-1)^\kappa\vec{n}\cdot\vec{\sigma}]/2$. $I$ represents the identity matrix and $\vec{\sigma}=(\sigma_x,\sigma_y,\sigma_z)$ is the Pauli vector. All of the CSs form an assemblage on Bob, as introduced in detail in Ref. \cite{piani2015}. It is worth mentioning that Bob's unconditional state $\rho_B=\textup{Tr}_A[\rho_{AB}]=\sum_\kappa\tilde{\rho}_{\kappa|\vec{n}}$ remains unchanged regardless of the measurement direction $\vec{n}$ Alice chooses. After obtaining all the CSs, Bob can judge whether there exists an LHSM consisting of the state ensemble of $E_{LHSM}=\{p_i\rho_i\}$ satisfying $\rho_B=\sum_ip_i\rho_i$, where $\rho_i$ is the normalized local hidden state with the corresponding probability $p_i$. $\sum_ip_i=1$ and $p_i\in[0,1]$. For the case of two measurement settings, it has been proven that four local hidden states are sufficient to reproduce the four CSs if an LHSM exists \cite{wcf2014}. If Bob's CSs can be rewritten as the combinations of $E_{LHSM}$ shown below

\begin{equation}\label{rho-lhsm}
\tilde{\rho}_{\kappa|\vec{n}}=\sum_i P(\kappa|\vec{n},i)p_i\rho_i,
\end{equation} then the steering task fails. The probability distribution $P(\kappa|\vec{n},i)$ is a stochastic map (positive and normalized) from $i$ to $\kappa$. As proven in Ref. \cite{wcf2014}, for any given two-qubit state, $P(\kappa|\vec{n},i)\in\{0,1\}$ for all the $\kappa$, $\vec{n}$ and $i$. The simulation of Bob's four CSs is demonstrated through the local hidden states channel, as shown in Fig. \ref{theory}, and the corresponding equations derived based on Eq. (\ref{rho-lhsm}) can be written as follows:

\begin{equation}\label{lhs}
\left\{
\begin{aligned}
t_{C1} &= p_a + p_d \texttt{;} \;\; \tilde{\rho}_{C1} = p_a \rho_a + p_d \rho_d \\
t_{C2} &= p_b + p_c \texttt{;} \;\; \tilde{\rho}_{C2} = p_b \rho_b + p_c \rho_c \\
t_{D1} &= p_c + p_a \texttt{;} \;\; \tilde{\rho}_{D1} = p_c \rho_c + p_a \rho_a \\
t_{D2} &= p_d + p_b \texttt{;} \;\; \tilde{\rho}_{D2} = p_d \rho_d + p_b \rho_b
\end{aligned}
\right.
\end{equation} where $t_i=\textup{Tr}[\tilde{\rho}_i]$. Each CS can now be reproduced by a combination of only two elements from the ensemble $E_{LHSM}$. Otherwise, if there is no such ensemble $E_{LHSM}=\{p_i\rho_i\}$ and the map distribution $P(\kappa|\vec{n},i)$ satisfying Eq. (\ref{lhs}), Bob confirms that Alice steers his system successfully.

We can expand the hidden states $\rho_i$ ($i=a$, $b$, $c$, $d$) to the super quantum hidden state model (SQHSM), which means there are no physical restrictions on the states $\rho_i$ and $\rho_i$, which can be located outside of the Bloch sphere. In such a case, there is generally more than one set of solutions of the linear equations Eq. (\ref{lhs}). Employing the quantum steering ellipsoids \cite{jevtic2014}, for any given two-qubit state $\rho_{AB}$ and the set of two measurement settings $\{\vec{n}_1,\vec{n}_2\}$, the radius of the SQHSM is defined as
 \begin{equation}\label{sr}
 r(\rho_{AB})_{\{\vec{n}_1,\vec{n}_2\}}=\min_{SQHSM}\{\max\{L[\rho_a],L[\rho_b],L[\rho_c],L[\rho_d]\}\},
 \end{equation}
 where $L[\rho_i]$ ($i=a$, $b$, $c$, $d$) denotes the length of Bloch vectors of the states $\rho_i$. If $r(\rho_{AB})_{\{\vec{n}_1,\vec{n}_2\}}>1$, at least one of the hidden states is located beyond the Bloch sphere; thus, $E_{LHSM}$ is not a physical ensemble. Because we can choose any measurement settings to demonstrate the steerability, the steering radius is defined as
 \begin{equation}\label{steerrad}
 R(\rho_{AB})=\max_{\{\vec{n}_1,\vec{n}_2\}}\{r(\rho_{AB})_{\{\vec{n}_1,\vec{n}_2\}}\}.
 \end{equation}
 In a recent work \cite{piani2015}, the steering robustness was defined to quantify the steerability for the state $\rho_{AB}$. The steering robustness is defined as $\mathcal{R}(\mathcal{A}):=\min\{t\geq0|\{\Xi_{a|x}\}_{a,x}\}$, where $\mathcal{A}=\{\tilde{\rho}_{a|x}\}_{a,x}$ is an assemblage given by $\rho_{AB}$, and $\Xi_{a|x}=(\tilde{\rho}_{a|x}+t\tilde{\tau}_{a|x})/(1+t)$ is unsteerable with an arbitrary assemblage $\{\tilde{\tau}_{a|x}\}_{a,x}$. If we pick $\tilde{\tau}_{a|x}=\textup{Tr}[\tilde{\rho}_{a|x}]\cdot I/2$, and restrict the number of measurement settings to two, we find that $R(\rho_{AB})=1+\mathcal{R}(\mathcal{A})$. As a result, $R(\rho_{AB})>1$ is the necessary and sufficient condition that the state $\rho_{AB}$ is steerable in the two measurement settings case.

Asymmetry is an inherent characteristic of EPR steering. We consider the two-qubit asymmetric states as follows,
\begin{equation}\label{rho1}
\rho_{AB}=\eta|\Psi(\theta)\rangle\langle\Psi(\theta)|+(1-\eta)|\Phi(\theta)\rangle\langle\Phi(\theta)|,
\end{equation}
with $0\leq{\eta}\leq 1$. $|\Psi(\theta)\rangle=\cos{\theta}|0_A 0_B\rangle+\sin{\theta}|1_A 1_B\rangle$ and $|\Phi(\theta)\rangle=\cos{\theta}|1_A 0_B\rangle+\sin{\theta}|0_A 1_B\rangle$. Here, $|0_A\rangle$ and $|1_A\rangle$ ($|0_B\rangle$ and $|1_B\rangle$) are the computational basis of Alice's (Bob's) qubit. It has been proven that, for all non-trivial states $\rho_{AB}$ (entangled), Alice can always steer Bob's system in the case of two measurement settings along directions $x$ and $z$ \cite{chenjl2013}. Referring to the steering radius, we find $R(\rho_{AB})=r(\rho_{AB})_{\{x,\,z\}}$. As a result, $R(\rho_{AB})=r(\rho_{AB})_{\{x,\,z\}}>1$ for all non-trivial $\rho_{AB}$. Considering the steering process from Bob to Alice, we prove that Bob could $not$ steer Alice's state when $|\cos 2\theta|\geq |2\eta -1|$ \cite{sm}. Therefore, if the state $\rho_{AB}$, which Alice and Bob shared, is located in the area satisfying $|\cos 2\theta|\geq |2\eta -1|$, there always exists an LHSM for Alice to reproduce her CSs when Bob chooses any two directions to measure. Under such a condition, the state $\rho_{AB}$ possesses the property of one-way steering. If there exists an LHSM for Alice when Bob measures along $x$ and $z$, then $|\cos2\theta|\geq |2\eta-1|$ \cite{sm}. As a result, we can focus only on the measurement directions $\{x,\,z\}$ in demonstrating the two-measurement-setting steering protocol. For the non-steerability case, the four elements of $E_{LHSM}$ are located in the $XZ$ plane of the Bloch sphere. We can then prepare the local hidden states with the corresponding probabilities severally. When $|\cos 2\theta| < |2\eta-1|$, we prove that $R(\rho_{BA})=r(\rho_{BA})_{\{x,z\}}$ and find that $R(\rho_{AB})>R(\rho_{BA})>1$, which shows that the asymmetry still exists in the case of two-way steering \cite{sm}.

\begin{figure}
\centering
\includegraphics[width=0.47\textwidth]{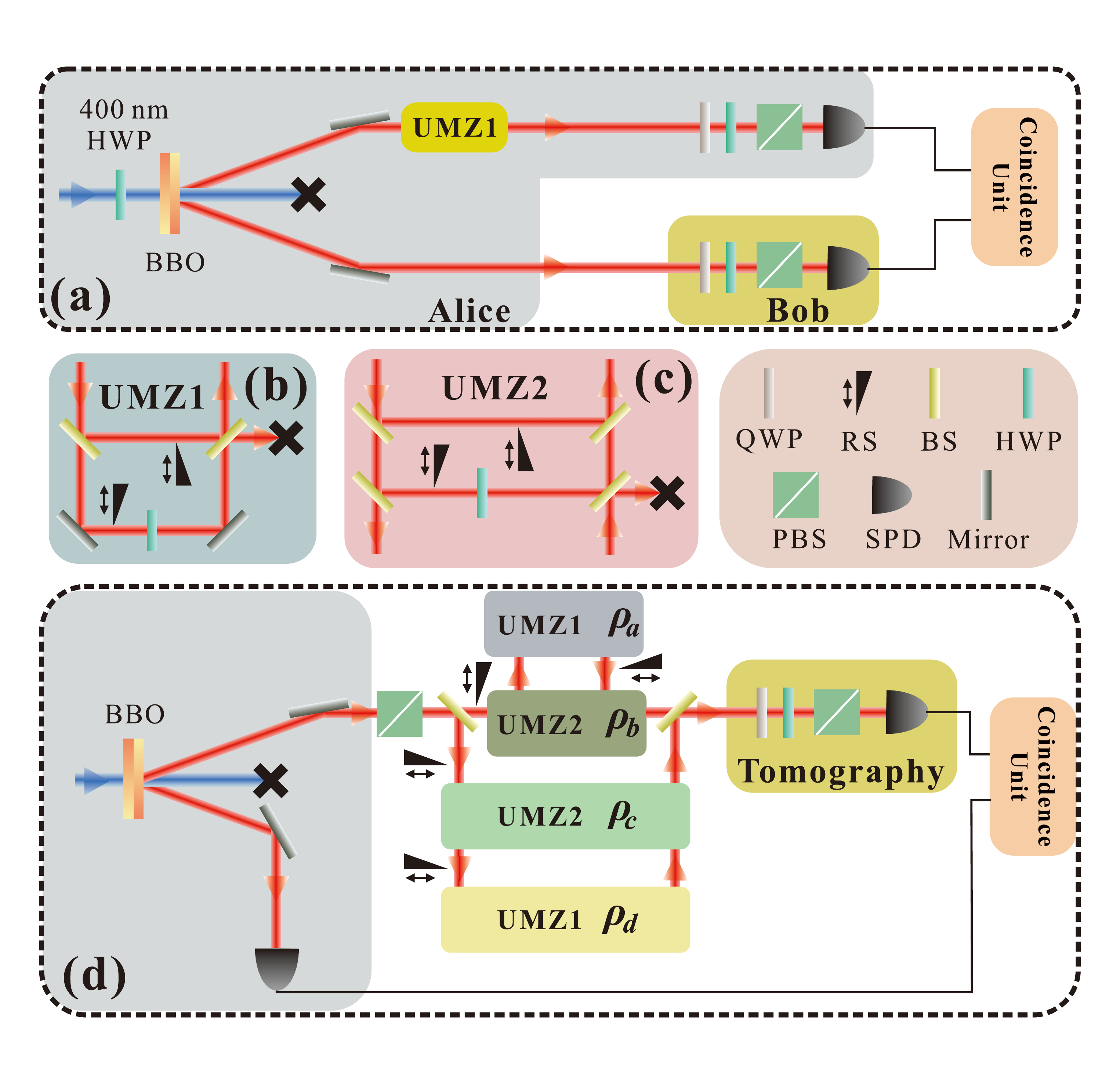}
\caption{Experimental setup. {\bf (a)}. The entangled photon pairs are prepared through the spontaneous parametric down conversion (SPDC) process by pumping the BBO crystal with ultraviolet pulses. The state's parameters $\eta$ and $\theta$ can be detuned conveniently by employing the setup shown in {\bf (a)} and the unbalanced Mach-Zehnder interferometer (UMZ) with beam splitters (BSs) and removable shutters (RSs) shown in {\bf (b)}. A unit consisting of a quarter-wave plate (QWP) and a half-wave plate (HWP) on Alice's side is used to set the measurement direction. The same unit with an extra polarization beam splitter (PBS) on Bob's side is used to perform state tomography. Photons are collected into a single mode fiber equipped with a 3 $nm$ interference filter and are then detected by a single-photon detector (SPD) on each side. {\bf (d)}. The strategy is for $E_{LHSM}$ to reproduce the CSs. One of the two photons is used as the trigger for the coincidence unit, and the other is used to prepare the four local hidden states, which can be conveniently prepared by employing the setup of {\bf (b)} and {\bf (c)}. The probabilities are controlled by adjusting the RSs.
}
\label{setup}
\end{figure}

Fig. \ref{setup} {\bf (a)} and {\bf (b)} show the setup for preparing the asymmetric entangled states in Eq. (\ref{rho1}). Ultraviolet pulses with a center wavelength of 400 $nm$ and a bandwidth of approximately 1.2 $nm$ are used to pump the two-crystal geometry type-I BBO crystals to generate entangled photon pairs \cite{kwiat1999}. A 400 $nm$ half-wave plate (HWP) is used to change the state's parameter $\theta$. When we consider the steering process from Alice to Bob, Bob calculates the steering radius $R$ with the four obtained CSs and determines whether the steering is successful. If an LHSM exists, we further check this fact by constructing the ensemble $E_{LHSM}$ with the UMZ inserted with an HWP and two RSs, as shown in Fig. \ref{setup} {\bf (b)} and {\bf (c)}. Following the corresponding combinational rules of hidden states, Bob can reproduce the four CSs, as shown in Fig. \ref{setup} {\bf (d)}. In contrast, when Bob wants to steer Alice's system, Bob measures his qubit along the directions $x$ and $z$, and Alice performs state tomography of her CSs. To verify the existence of an LHSM, Alice attempts to reconstruct the CSs using four local hidden states with the same setup shown in Fig. \ref{setup} {\bf (d)}.

\begin{figure}
\centering
\includegraphics[width=0.47\textwidth]{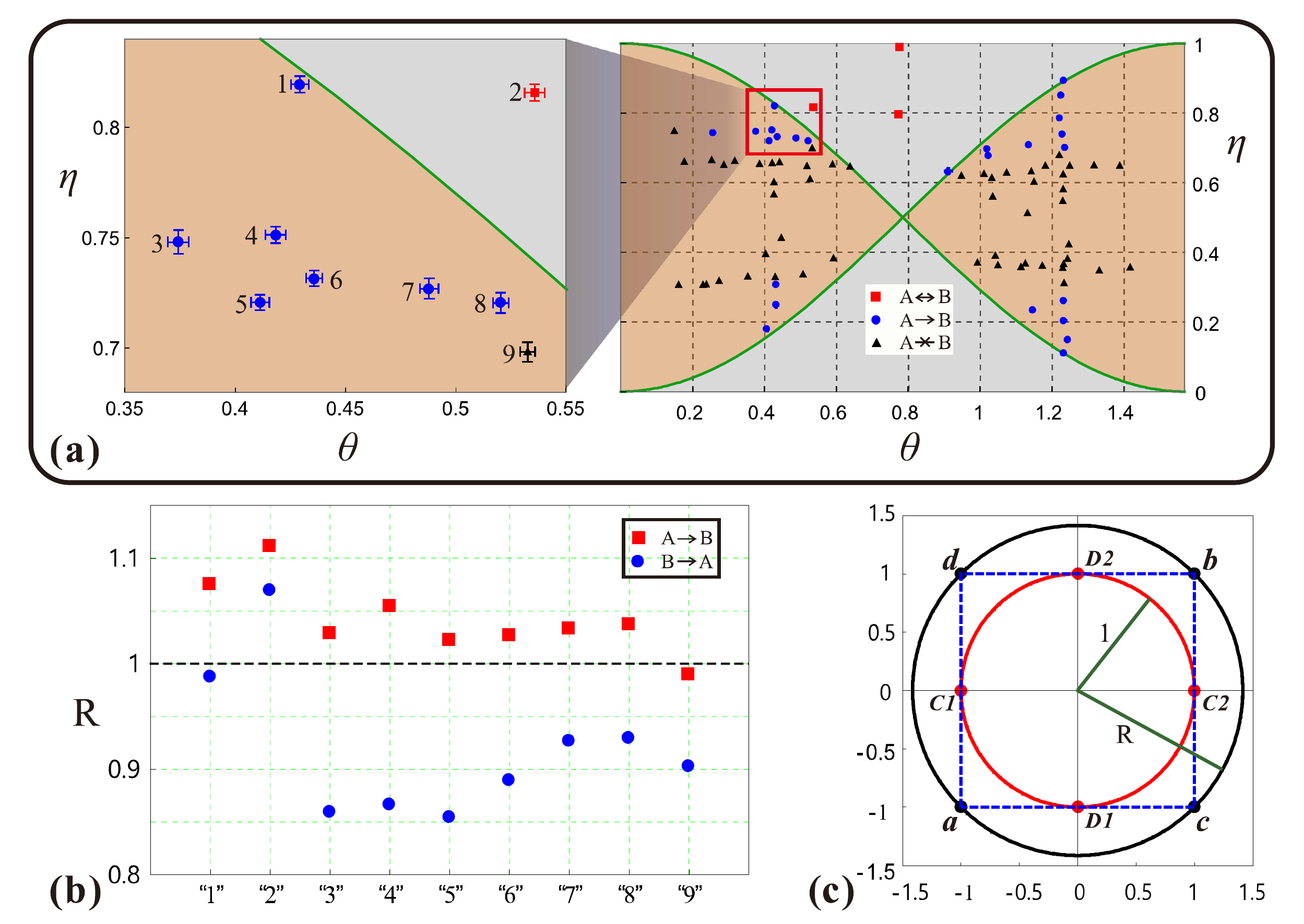}
\caption{Experimental results for asymmetric EPR steering. {\bf (a).} The distribution of the experimental states. The right column shows the entangled states we prepared, and the left column is a magnification of the corresponding region in the right column. The two green curves represent the cases of $|\cos 2\theta| = |2\eta -1|$. The blue points and red squares represent the states realizing one-way and two-way EPR steering, respectively. The black triangles represent the states for which EPR steering task fails for both observers. {\bf (b).} The values of $R$ for the states labeled in the left column in {\bf (a)}. The red squares represent the situation where Alice steers Bob's system, and the blue points represent the case where Bob steers Alice's system. {\bf (c).} Geometric illustration of the strategy for local hidden states (black points) to construct the four normalized CSs (Red points) obtained from the maximally entangled state.}
\label{result1}
\end{figure}

We prepare several entangled states in the form of $\rho_{AB}$ to perform the EPR steering task. In our experiment, Alice (Bob) can obtain the normalized CSs with fidelities of approximately $99.3 \pm 0.3\%$. In Fig. \ref{result1} {\bf (a)}, the two yellow regions satisfy $|\cos 2\theta|\geq |2\eta -1|$. According to previous theoretical analysis, Bob cannot steer Alice's state when the shared states are located in these two regions. However, owing to the coordinate errors of the experimental CSs when represented in the Bloch sphere \cite{sm}, which are deduced from the counting statistics, several states, for which Alice can steer Bob in theory cannot be used to complete the EPR steering task for both Alice and Bob with the measurement directions along $x$ and $z$. The states represented by blue points show the case of one-way steering, i.e., Alice can steer Bob's state $(A\rightarrow B)$, but Bob could $not$ steer Alice's state. The red squares represent the states that show the cases for two-way steering (Alice and Bob can steer each other $(A\leftrightarrow B)$). Asymmetry still exists in such a case because the values of $R(\rho_{AB})$ are larger than the corresponding values of $R(\rho_{BA})$ in both theory and experiment \cite{sm}. A small region, which is surrounded by the red frame in the right column, is magnified and shown in the left column, where the states are labeled by numbers. The corresponding values of $R$ are shown in Fig. \ref{result1} {\bf (b)}, which demonstrates the asymmetry of steering. The EPR steering task is successful if $R > 1$. Otherwise, the EPR steering fails. The strategy to use local hidden states (black points) to construct the four normalized CSs (red points) obtained from the maximally entangled state is shown in Fig. \ref{result1}. {\bf (c)}.

\begin{figure}
\centering
\includegraphics[width=0.47\textwidth]{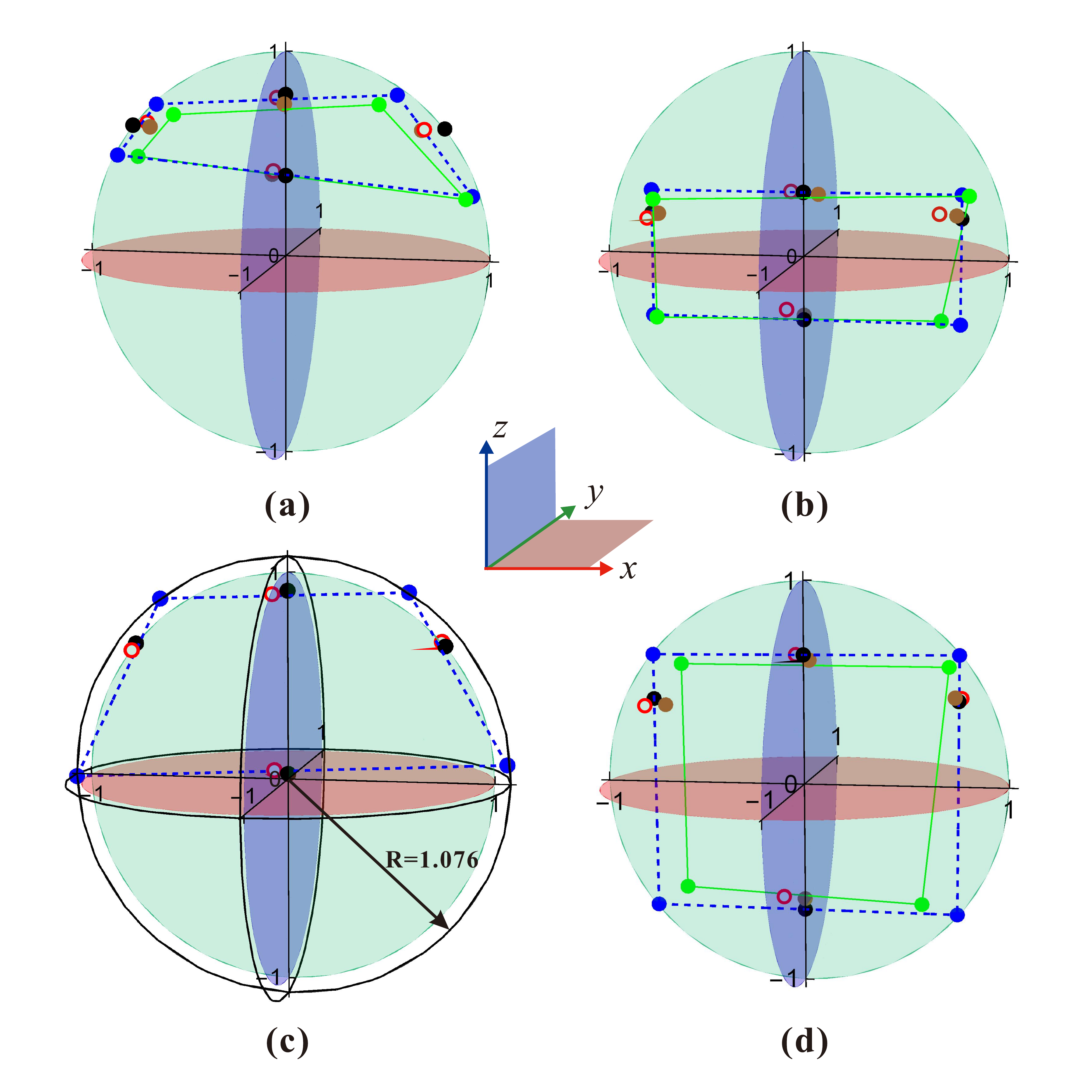}
\caption{The experimental results of the normalized CSs and local hidden states shown in the Bloch sphere. The theoretical and experimental results of the normalized CSs are marked by the black and red points (hollow), respectively. The blue and green points represent the results of the four local hidden states in theory and experiment, respectively. The normalized CSs constructed by the local hidden states are shown by the brown points. {\bf (a)} and {\bf (c)} show the case in which Alice steers Bob's system, whereas {\bf (b)} and {\bf (d)} show the case in which Bob steers Alice's system. The parameters of the shared state in {\bf (a)} and {\bf (b)} are $\theta=0.442$ and $\eta=0.658$; the parameters of the shared state in {\bf (c)} and {\bf (d)} are $\theta=0.429$ and $\eta=0.819$. {\bf (a)}, {\bf (b)} and {\bf (d)} show that the LHSMs exist, and the steering tasks fail. {\bf (c)} shows that no LHSM exists for the steering process with the constructed hidden states located beyond the Bloch sphere and $R=1.076$.}
\label{result2}
\end{figure}

In our experiment, we construct the corresponding local hidden states if an LHSM exists. For the states that were theoretically predicted to show the ability of EPR steering but failed because of experimental errors, the ensemble $E_{LHSM}$ is deduced based on the experimental results of the CSs. For the states that could not be used to realize the EPR steering theoretically, we use the theoretically predicted CSs to deduce the ensemble. Fig. \ref{result2} shows the experimental combinations of the local hidden states to reproduce CSs. The constructed CSs are obtained with high fidelity of approximately $99.8\pm0.1\%$ compared to the desired states. To illustrate the one-way EPR steering, a failure to construct the hidden states is shown in Fig. \ref{result2}. {\bf (c)} where the theoretical hidden states are located outsides the Bloch sphere. There exists a situation in which only three hidden states are sufficient to rebuild the four CSs, and the results are shown in the Supplementary material \cite{sm}. Additional results with the measurement directions set to be $\{x,\,y\}$ and $\{y,\,z\}$ can be found in the Supplementary material. The experimental errors are estimated from the statistical variation of photon counts, which satisfy the Poisson distribution.

The LHSM provides a direct and convinced contradiction between the nonlocal EPR steering and classical physics. In this work, we propose a feasible way to find the LHSM for the case where EPR steering fails. Following a similar idea to steering robustness \cite{piani2015}, we introduce a practical criterion $R$ to quantify the steerability of entangled states in the case of two settings measurement. The experimental results show the asymmetry of EPR steering and the presence of one-way steering with projective measurements. We further experimentally prepare the local hidden states and use them to reconstruct the CSs with high fidelity when $R\leq1$.

Our protocol is restricted to two-setting measurement. Any quantum information application or experimental demonstration is realized by a finite measurement setting scenario. Taking the semi-device independeded quantum key distribution as an example \cite{wiseman2012pra}, Alice and Bob will have a certain finite setting protocol (e.g., two-setting) before the steering resource is distributed. Once the steerability is demonstrated by this two-setting protocol, the security of the key is not threatened by the number of measurement settings. Our work provides an intuitional and fundamental way to understand the EPR steering and the asymmetric nonlocality. The demonstrated asymmetric EPR steering has important applications in the tasks of one-way quantum key distribution \cite{gisin2002} and the quantum subchannel discrimination \cite{piani2015}, even within the frame of two-setting measurements.

\noindent\emph{Note added}: Another experiment \cite{wollmann2015} is performed to realize one-way EPR steering using the Werner state with a lossy channel at one side.

\newpage{}

\section{Appendix}

\subsection{One-way EPR steering for state $\rho_{AB}$}

\begin{figure}[h]
\includegraphics[width=0.4\textwidth]{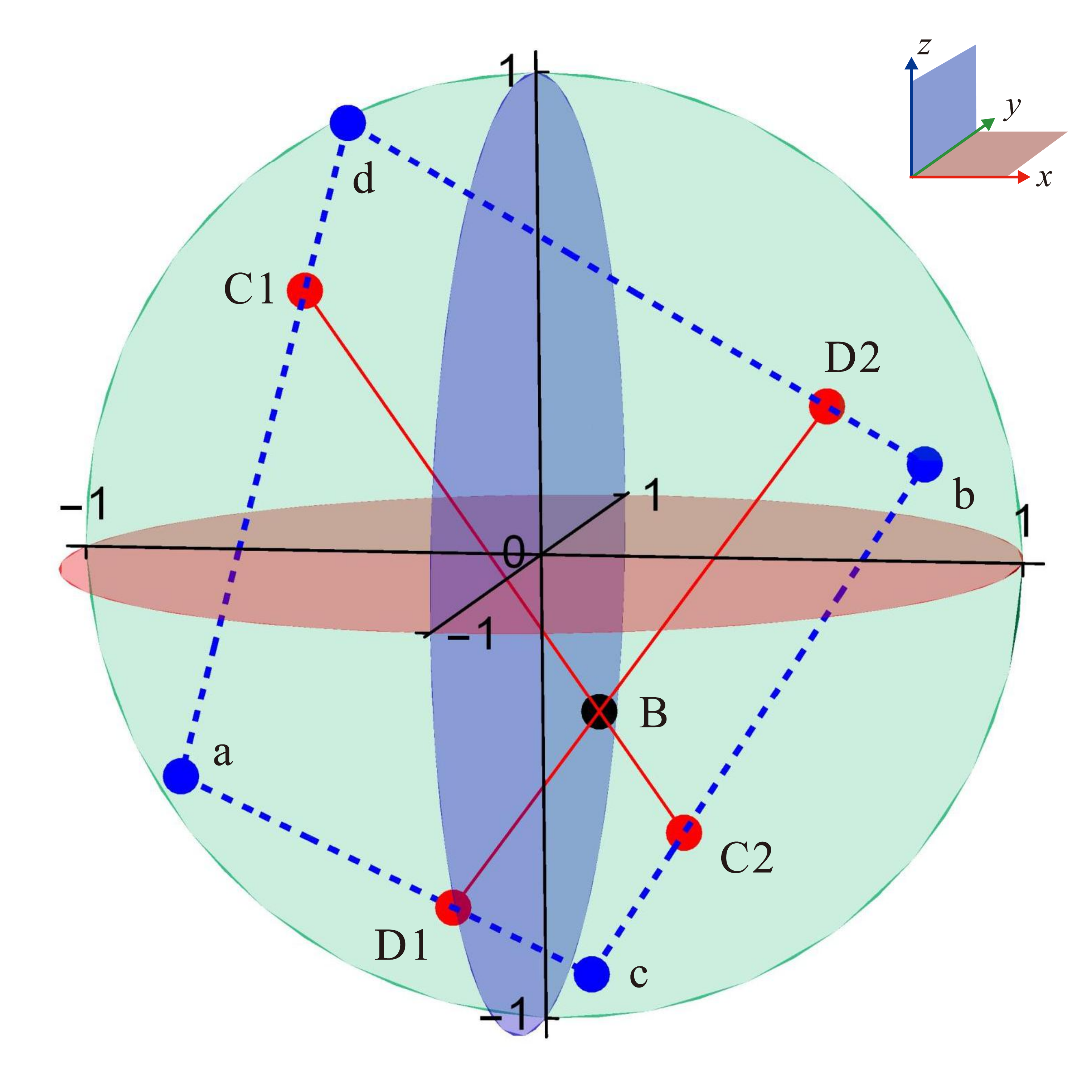}\\
\caption{Four hidden states simulating the four normalized conditional states in the Bloch sphere. The four blue points, labeled as $a, b, c$ and $d$, are the hidden states, and the four red points, $C1,\,C2,\,D1$ and $D2$, stand for the normalized conditional states. The black point $B$ represents the normalized unconditional state. It is shown that, $\{a,\,d\}$ simulate $C1$, $\{b,\,c\}$ simulate $C2$, $\{c,\,a\}$ simulate $D1$ and $\{d,\,b\}$ simulate $D2$.}\label{lhs-cs}
\end{figure}

As mentioned in the main text, for the state $\rho_{AB}$, Alice could always steer Bob's system except for some non-trivial states, which has been proved in Ref. \cite{wcf2014,chenjl2013}. Bob could $not$ steer Alice's state when $|\cos2\theta|\geq|2\eta-1|$ in the case of two measurement settings. The one-way EPR steering process can be observed when $|\cos2\theta|\geq|2\eta-1|$. Here, the local hidden state model (LHSM) for Alice is directly given to prove this point.

The four local hidden states shown in Fig. \ref{lhs-cs} are represented as $a=(a_x,a_y,a_z)$, $b=(b_x,b_y,b_z)$, $c=(c_x,c_y,c_z)$ and $d=(d_x,d_y,d_z)$. The states are located in the Bloch sphere, and the corresponding normalized conditional states are represented as $C1$, $C2$, $D1$ and $D2$. From the geometry relationship between them, we can obtain

\begin{equation}\label{one-waylhs1}
\left\{ \begin{aligned}
         a_x &= \frac{C1_x+D1_x}{1+C1_xD1_x} \\
         a_y &= \frac{C1_y D1_z+C1_z D1_y}{(1+C1_xD1_x)\cos2\theta(2\eta-1)} \\
         a_z &= \frac{C1_z D1_z-C1_y D1_y}{(1+C1_xD1_x)\cos2\theta(2\eta-1)}
        \end{aligned} \right. \tag{3.a}
\end{equation}
\begin{equation}\label{one-waylhs2}
 \left\{ \begin{aligned}
         b_x &= \frac{C2_x+D2_x}{1+C2_xD2_x} \\
         b_y &= \frac{C2_y D2_z+C2_z D2_y}{(1+C2_xD2_x)\cos2\theta(2\eta-1)} \\
         b_z &= \frac{C2_z D2_z-C2_y D2_y}{(1+C2_xD2_x)\cos2\theta(2\eta-1)}
        \end{aligned} \right. \tag{3.b}
  \end{equation}
  \begin{equation}\label{one-waylhs3}
\left\{ \begin{aligned}
         c_x &= \frac{C2_x+D1_x}{1+C2_xD1_x} \\
         c_y &= \frac{C2_y D1_z+C2_z D1_y}{(1+C2_xD1_x)\cos2\theta(2\eta-1)} \\
         c_z &= \frac{C2_z D1_z-C2_y D1_y}{(1+C2_xD1_x)\cos2\theta(2\eta-1)}
        \end{aligned} \right. \tag{3.c}
\end{equation}
\begin{equation}\label{one-waylhs4}
 \left\{ \begin{aligned}
         d_x &= \frac{C1_x+D2_x}{1+C1_xD2_x} \\
         d_y &= \frac{C1_y D2_z+C1_z D2_y}{(1+C1_xD2_x)\cos2\theta(2\eta-1)} \\
         d_z &= \frac{C1_z D2_z-C1_y D2_y}{(1+C1_xD2_x)\cos2\theta(2\eta-1)}
        \end{aligned} \right. \tag{3.d}
 \end{equation}
with the probabilities

\begin{equation}\label{one-waypro}
 \left\{ \begin{aligned}
         p_a &= \frac{|BD2|}{|D1D2|}\frac{|cD1|}{|ac|}\\
         p_b &= \frac{|BD1|}{|D1D2|}\frac{|dD2|}{|bd|}\\
         p_c &= \frac{|BD2|}{|D1D2|}\frac{|aD1|}{|ac|}\\
         p_d &= \frac{|BD1|}{|D1D2|}\frac{|bD2|}{|bd|}\\
        \end{aligned} \right.
\end{equation} where $|D1D2|$ is the distance between point $D1$ and $D2$. The points of four normalized conditional states, i.e., $\{C1,\,C2,\,D1,\,D2\}$, are located on the ellipsoid $x^2+\frac{y^2}{{r_e}^2}+\frac{z^2}{{r_e}^2}=1$ where $r_e=2\eta-1$.

To prove the above ensemble satisfying the equations (1) mentioned in the main text, we demonstrate that the four constricted states are first physical states. By defining $R_e=|\frac{2\eta-1}{\cos2\theta}|$, we obtain ${a_x}^2+\frac{{a_y}^2}{{R_e}^2}+\frac{{a_z}^2}{{R_e}^2}=1$, which means the point $a$ is on an ellipsoid. Following a similar method, we deduce that $b$, $c$ and $d$ are all on ellipsoids. These points are not beyond Bloch sphere when $|\cos2\theta|\geq|2\eta-1|$. For $p_i$ ($i=a,b,c,d$), we obtain $\sum_ip_i=1$ and $p_i\in[0,1]$. By direct calculation, equations (1) can be obtained.

In contrast, if there exists such an LHSM for Alice when Bob measures along $x$ and $z$, we can know that the shared state is located at the regions with $|\cos2\theta|\geq|2\eta-1|$. As a result, Alice could always find an LHSM, regardless of which two directions Bob chooses to measure.

The above discussion are valid for general cases. There is a singular situation where the denominators of Eqs. (3) are $0$. For the four states ($a,b,c$ and $d$), only one can be singular. Let us assume this state is $a$. Thus, $C1_xD1_x=-1$. So, $C1_x=-1(1)$, $D1_x=1(-1)$ and the other coordinates of $C1$ and $D1$ are all $0$. The hidden states are only $b$, $c$ and $d$. We obtain that $b=(0,0,\frac{C2_zD2_z}{(1+C2_xD2_x)}\frac{1}{\cos2\theta(2\eta-1)})=\frac{2\eta-1}{\cos2\theta}$, $c=D1$ and $d=C1$ with the probabilities $p_c=t_{D1}$, $p_d=t_{C1}$ and $p_b=1-p_c-p_d$.

\subsection{The analysis of the steering radius}

Steering radius provides an intuitive insight into presenting the contradiction between EPR steering and classical physics in the Bloch sphere. It can be employed in the situation where the conditional states (CSs) can be shown in the Bloch sphere. As introduced in the main text, there is a simple relation between steering radius and steering robustness in two-qubit system, which is $R(\rho_{AB})=1+\mathcal{R}(\mathcal{A})$ if one picks $\tilde{\tau}_{a|x}=\textup{Tr}[\tilde{\rho}_{a|x}]\cdot I/2$. The proof is shown as follows.

By setting the assemblage $\tilde{\tau}_{a|x}$ as $\textup{Tr}[\tilde{\rho}_{a|x}]\cdot I/2$, the unsteerable assemblage $\Xi_{a|x}$ becomes $\Xi_{a|x}=(\tilde{\rho}_{a|x}+t\textup{Tr}[\tilde{\rho}_{a|x}]\cdot I/2)/(1+t)$. There is an LHSM with $\{p_i\rho_i\}\, (i=a,\,b,\,c,\,d)$ for $\Xi_{a|x}$. Then the Eqs. (2) in the main text can be rewritten as
\begin{widetext}
\begin{equation}\label{lhss}
\left\{
\begin{aligned}
t_{C1} &= p_a + p_d \texttt{;} \;\; \frac{1}{1+t}t_{C1}\rho_{C1}+\frac{t}{1+t}t_{C1}\cdot I/2 = p_a \rho_a + p_d \rho_d, \\
t_{C2} &= p_b + p_c \texttt{;} \;\; \frac{1}{1+t}t_{C2}\rho_{C2}+\frac{t}{1+t}t_{C2}\cdot I/2 = p_b \rho_b + p_c \rho_c, \\
t_{D1} &= p_c + p_a \texttt{;} \;\; \frac{1}{1+t}t_{D1}\rho_{D1}+\frac{t}{1+t}t_{D1}\cdot I/2 = p_c \rho_c + p_a \rho_a, \\
t_{D2} &= p_d + p_b \texttt{;} \;\; \frac{1}{1+t}t_{D2}\rho_{D2}+\frac{t}{1+t}t_{D2}\cdot I/2 = p_d \rho_d + p_b \rho_b.
\end{aligned}
\right.
\end{equation}
\end{widetext}
Rewriting the state $\rho_i$ by its Bloch vector $\vec{r}_i$, we can get
\begin{widetext}
\begin{equation}\label{lhss1}
\left\{
\begin{aligned}
t_{C1} &= p_a + p_d \texttt{;} \;\; t_{C1}\rho_{C1}=p_a \frac{I+(1+t)\vec{r}_a\cdot\vec{\sigma}}{2} + p_d \frac{I+(1+t)\vec{r}_d\cdot\vec{\sigma}}{2}, \\
t_{C2} &= p_b + p_c \texttt{;} \;\; t_{C2}\rho_{C2}=p_b \frac{I+(1+t)\vec{r}_b\cdot\vec{\sigma}}{2} + p_c \frac{I+(1+t)\vec{r}_c\cdot\vec{\sigma}}{2}, \\
t_{D1} &= p_c + p_a \texttt{;} \;\; t_{D1}\rho_{D1}=p_c \frac{I+(1+t)\vec{r}_c\cdot\vec{\sigma}}{2} + p_a \frac{I+(1+t)\vec{r}_a\cdot\vec{\sigma}}{2}, \\
t_{D2} &= p_d + p_b \texttt{;} \;\; t_{D2}\rho_{D2}=p_d \frac{I+(1+t)\vec{r}_d\cdot\vec{\sigma}}{2} + p_b \frac{I+(1+t)\vec{r}_b\cdot\vec{\sigma}}{2}.
\end{aligned}
\right.
\end{equation}
\end{widetext}
Let's define a super quantum hidden state model (SQHSM), $\{p_i^*\rho_i^*\}\,(i=a,\,b,\,c,\,d)$ with
\begin{equation}\label{lhss2}
p_i^*=p_i \texttt{;} \;\; \rho_i^*= \frac{I+\vec{L}_{i}\cdot\vec{\sigma}}{2}
\end{equation} where $\vec{L}_{i}=(1+t)\vec{r}_i$. It's easy to see $\{p_i^*\rho_i^*\}$ is an appropriate SQHSM for $\tilde{\rho}_{a|x}$. When $t$ takes the minimum value that allows $\Xi_{a|x}$ has a LHSM, $\vec{L}_{i}$ also takes the minimum value to construct a SQHSM for $\tilde{\rho}_{a|x}$. Thus, $R(\rho_{AB})=1+\mathcal{R}(\mathcal{A})$.

For the high-dimensional systems, the CSs on Bob's side don't have the Bloch vector expressions, thus the steering radius becomes invalid. The analogous method can be applied to higher-dimensional systems. The corresponding parameter could be defined as the minimum value $r$ which makes the assemblage $(\tilde{\rho}_{a|x}+(r-1)\textup{Tr}[\tilde{\rho}_{a|x}]\cdot I_N/N)/r$ has a local hidden state model (LHSM) description, where $N$ is the dimension on Bob's side and $I_N$ is $N\times N$ identity matrix. When $r_{min}>1$, Alice can steer Bob's state. However, the optimization method to obtain $r_{min}$ would require exponential increase of calculation resource in the case of higher-dimensional systems.

According to the detail definition of the steering radius, by the numerical analysis, $R(\rho_{AB})=r(\rho_{AB})_{\{x,z\}}$, and $R(\rho_{BA})=r(\rho_{BA})_{\{x,z\}}$ for the states satisfying $|\cos2\theta|<|2\eta-1|$. We further provide an analytic expression for $R(\rho_{AB})$ focusing on $\eta\in(1/2,1)$ and $\theta\in(0,\pi/2)$, and the expression is,
\begin{equation}\label{rab}
R(\rho_{AB})=\frac{1}{2\sqrt{2}(2\eta-1)}\sqrt{\frac{T1+T2}{1+\cos4\theta}},
\end{equation} where
\begin{equation}
T1=13-36\eta+36\eta^2+4(8\eta^2-8\eta+1)\cos4\theta-(1-2\eta)^2\cos8\theta
 \end{equation} and
\begin{widetext}
\begin{equation}
T2=\sqrt{2}(\cos4\theta-1)\sqrt{3-8\eta(\eta-1)(26\eta^2-26\eta+9)-4(1-2\eta)^2(12\eta^2-12\eta+1)\cos4\theta+(1-2\eta)^4\cos8\theta}.
\end{equation}
\end{widetext}

And when $|\cos2\theta|<|2\eta-1|$,
 \begin{equation}\label{rab}
R(\rho_{BA})=\sqrt{(2\eta-1)^2+(\sin2\theta)^2}.
\end{equation}

We can know that $R(\rho_{AB})>1$ for the non-trivial states and $R(\rho_{BA})>1$ when $|\cos2\theta|<|2\eta-1|$. Moreover $R(\rho_{AB})>R(\rho_{BA})>1$ when $|\cos2\theta|<|2\eta-1|$. Additionally, $R(\rho_{BA})=1$ when $|\cos2\theta|\geq|2\eta-1|$ according to the above discussion.

\begin{figure}[h]
\includegraphics[width=0.46\textwidth]{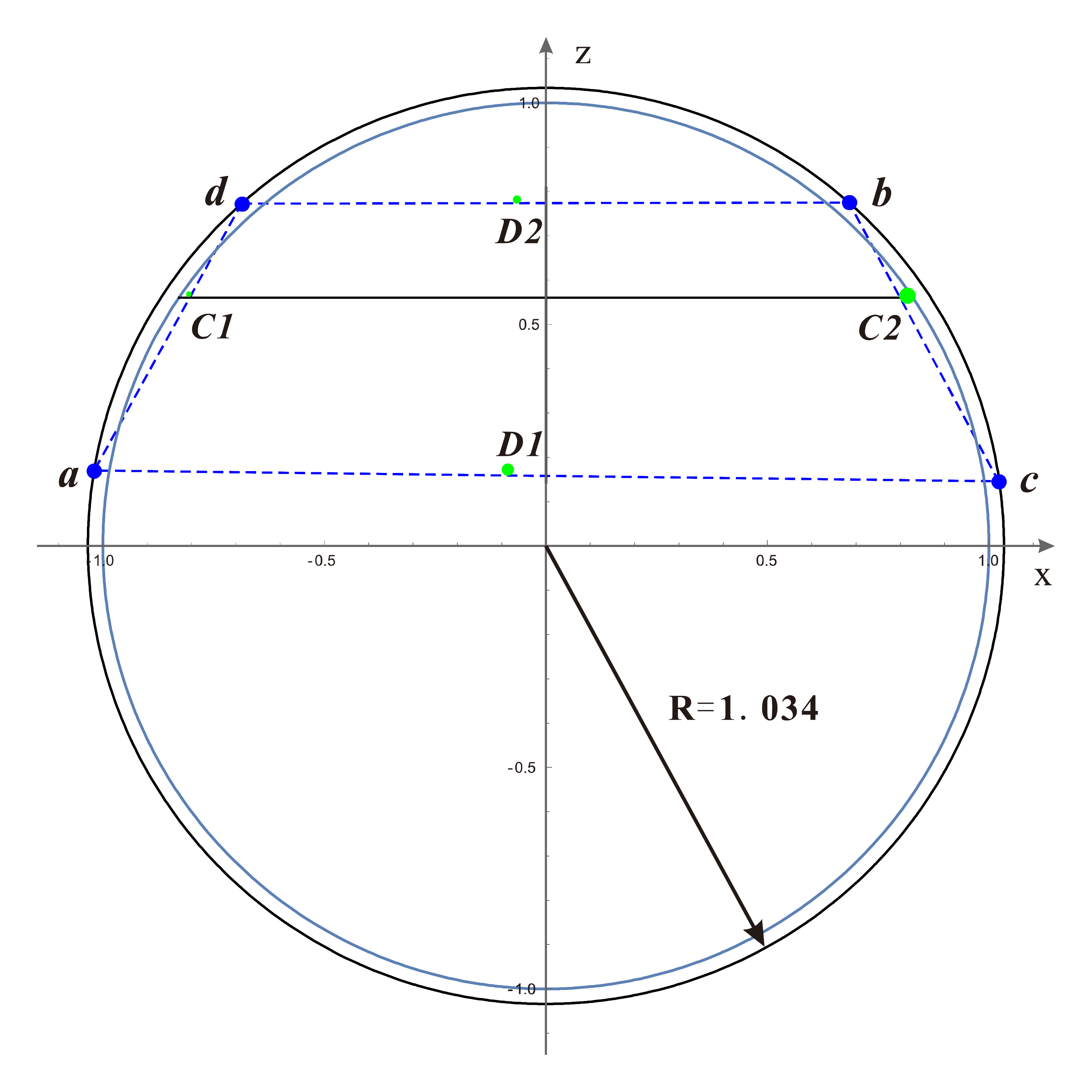}\\
\caption{The process of calculating the steering radius using the coordinate errors of Bob's experimental CSs in the Bloch sphere. The experimental results are obtained from the entangled state with $\theta=0.488$ and $\eta=0.727$. The blue points are the super quantum hidden states. The green circles, $C1$, $C2$, $D1$ and $D2$ with the corresponding radius of $0.0063, 0.0185, 0.0140$ and $0.0094$, represent the possible experimental CSs.}\label{figs11}
\end{figure}

In our experiment, considering the steering process from Alice to Bob, Bob could calculate the steering radius $R(\rho_{AB})$ by numerical optimization after obtaining the CSs $\tilde{\rho}_{C1},\, \tilde{\rho}_{C2}$ (Alice measures along $x$) and $\tilde{\rho}_{D1},\, \tilde{\rho}_{D2}$ (Alice measures along $z$). Because he doesn't trust either the source or Alice's measurement device, the calculation of $R(\rho_{AB})$ should only depend on his own measurement results. The normalized CSs can be represented in the Bloch sphere and the coordinates can be written as, saying the state $\rho_{C1}$, $C1=(x_{C1},\,y_{C1},\,z_{C1})$ with the corresponding errors $(Ex_{C1},\,Ey_{C1},\,Ez_{C1})$. Then the probable position of the state $\rho_{C1}$ in the Bloch sphere is within a sphere whose center is $C1$ and the error radius is $\sqrt{Ex_{C1}^2+Ey_{C1}^2+Ez_{C1}^2}$. To illustrate the calculative process clearly, $XZ$ plane of the Bloch sphere is presented in Fig. \ref{figs11}. By numerical optimization, Bob could find the minimum radius which contains all of the four super quantum hidden states (see the blue points). The minimization is over all possible SQHSM and all possible conditional states within the corresponding green circles. If this minimum radius $R(\rho_{AB})>1$, then Bob could certify that no LHSM can forge his measurement results and Alice can steer his state.

\subsection{Using the fidelity to demonstrate the one-way steering}

Generally, the probability of the experimentally determined state not being one-way steerable is low by virtue of its high fidelity with known one-way steerable states. However, states with the same fidelity calculated by $[Tr[\sqrt{\sqrt{\rho'}\rho_{theo}\sqrt{\rho'}}]]^2$ ($\rho'$ is the possible states and $\rho_{theo}$ is the theoretical prediction) may be quite different. It would be difficult to determine whether an experimental state is one-way steerable by only using its fidelity with the known one-way steerable states. Here, the requirement of fidelity of the source or the CSs to determine whether the states are one-way steerable is introduced.

{\bf $\bullet$ Demonstration of one-way steering by using the fidelity of the source}

\begin{figure}[h]
\includegraphics[width=0.45\textwidth]{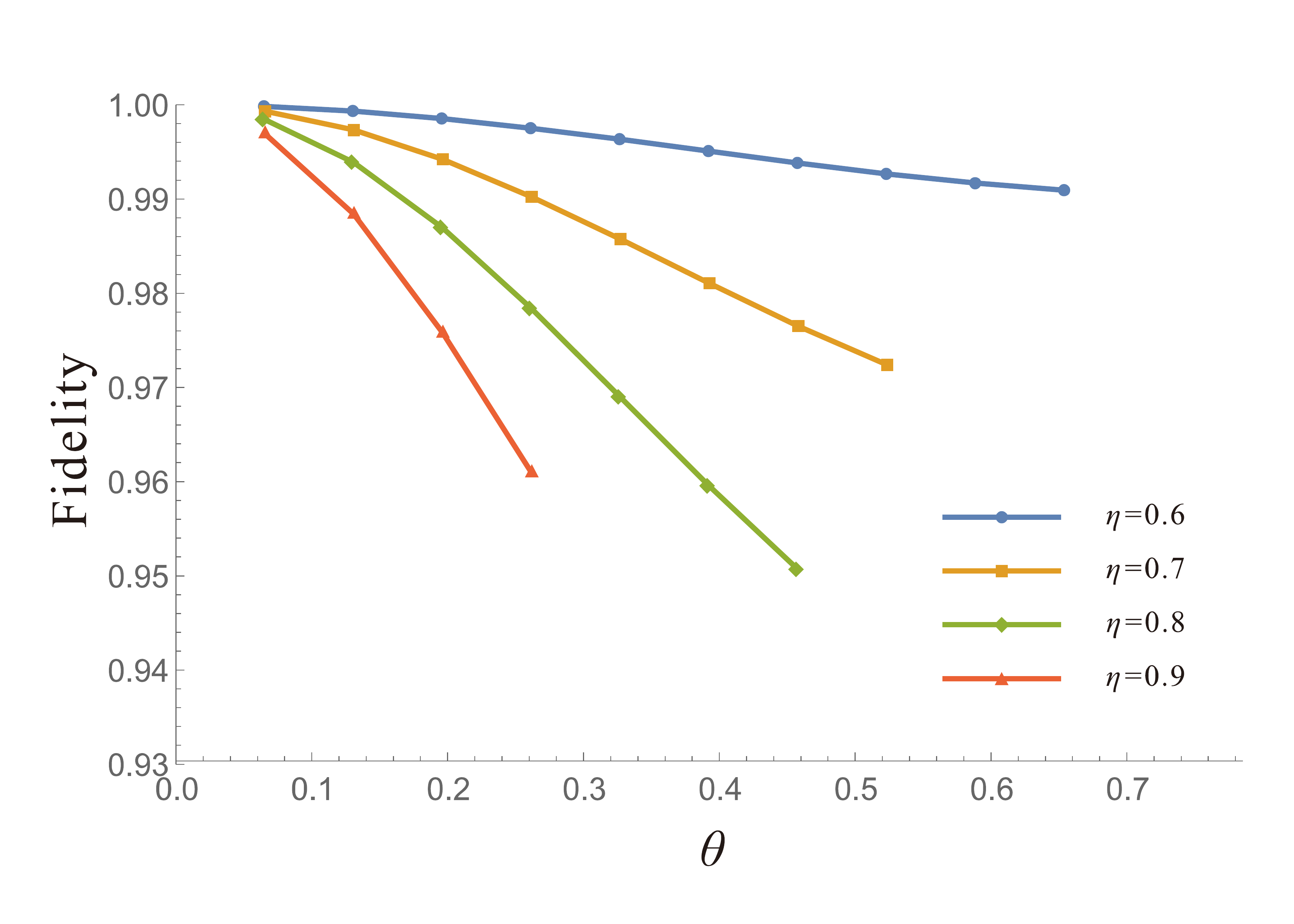}\\
\caption{The minimum fidelity of $\rho_{exp}$ to ensure that it is one-way steerable.}\label{figr2}
\end{figure}

There are always errors in experiments. For the sake of convenience, supposing the experimental errors could be described as the white noise, the original one-way steerable state ($\rho_{theo}$) turns to $\rho_{exp}=\lambda\rho_{theo}+(1-\lambda)I/4$ ($I$ is the identity matrix and $\lambda\in[0,1]$). Based on this noisy model, the fidelity of $\rho_{exp}$ can be used to determine whether the states are one-way steerable. Within the frame of our work, $\rho_{theo}$  locates in the area satisfying $|\cos 2\theta|\geq |2\eta -1|$. Because the steerability of $\rho_{exp}$ is decreased by the noise, it's easy to see that Bob still could not steer Alice's state by employing $\rho_{exp}$. On the other hand, the minimum $\lambda$ ensuring that Alice steers Bob's state with $\rho_{exp}$ can be calculated. Thus the minimum fidelity to determine the state $\rho_{exp}$ is one-way steerable is obtained. The detailed calculation is shown below.

Based on the method shown in the section V in the SM of Phys. Rev. Lett. 113, 140402 (2014), considering the symmetry of Bob's CSs, one could find the minimum $\lambda$ needed to demonstrate that Alice could steer Bob's state is:
\begin{widetext}
\begin{equation}\label{lambda}
\begin{split}
& \lambda_{min} =  \\
& \frac{4 [(1-2 \eta )^2 \cos 4\theta +2 \sqrt{(-2 (\eta -1) \eta -1) \sin ^4 2\theta [(1-2 \eta )^2 \cos 4\theta +4 (\eta -1) \eta -1]}+(1-2 \eta )^2]}{(1-2 \eta )^2 \cos 8\theta +28 (\eta -1) \eta -4 \cos 4\theta +11}.
\end{split}
\end{equation}
 \end{widetext}$\rho_{exp}$ is one-way steerable when $\lambda > \lambda_{min}$. The fidelity of $\rho_{exp}$ is minimized when $\lambda=\lambda_{min}$. Thus, the state $\rho_{exp}$ can be determined to be one-way steerable when its fidelity is higher than the corresponding one. The results of some states are illustrated in the Fig. \ref{figr2}.

It's worthy mentioning that the real state prepared in the experiment generally doesn't follow this white noise model. The above discussion of using the fidelity of the source to demonstrate one-way steering could only be considered as a guidance instead of a criterion.

{\bf $\bullet$ Using the fidelities of the CSs to ensure one-way steering}

\begin{figure}[h]
\includegraphics[width=0.45\textwidth]{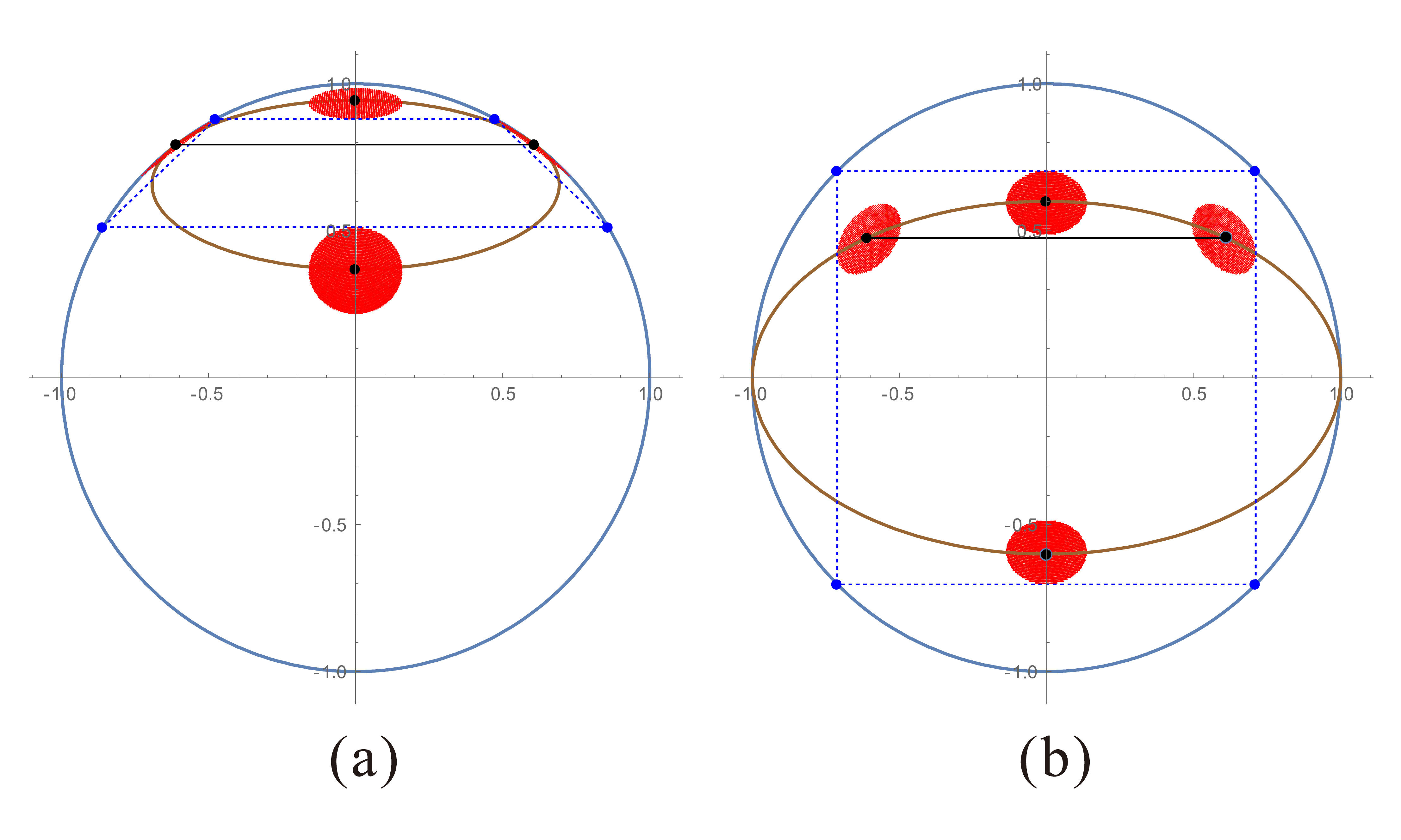}\\
\caption{Illustration of one-way steering using the fidelities of the CSs shown in the Bloch sphere. The brown ellipses are the conditional states' ellipses. The black points represent the theoretical CSs. The blue points represent the local hidden states. The four red regions represent the states whose fidelities are larger than the given value of fidelity. (a) Illustration of the process that Alice can steer Bob's state by using the minimum fidelity of Bob's four CSs. The given value of the fidelity is 0.9937 and $\theta=5\pi/48,\,\eta=0.8$. (b) Illustration of the process that Alice's state cannot be steered by Bob by using the minimum fidelity of Alice's four CSs. The given value of the fidelity is 0.9954 and $\theta=5\pi/48,\,\eta=0.8$.}\label{figr3}
\end{figure}

\begin{figure}[h]
\includegraphics[width=0.45\textwidth]{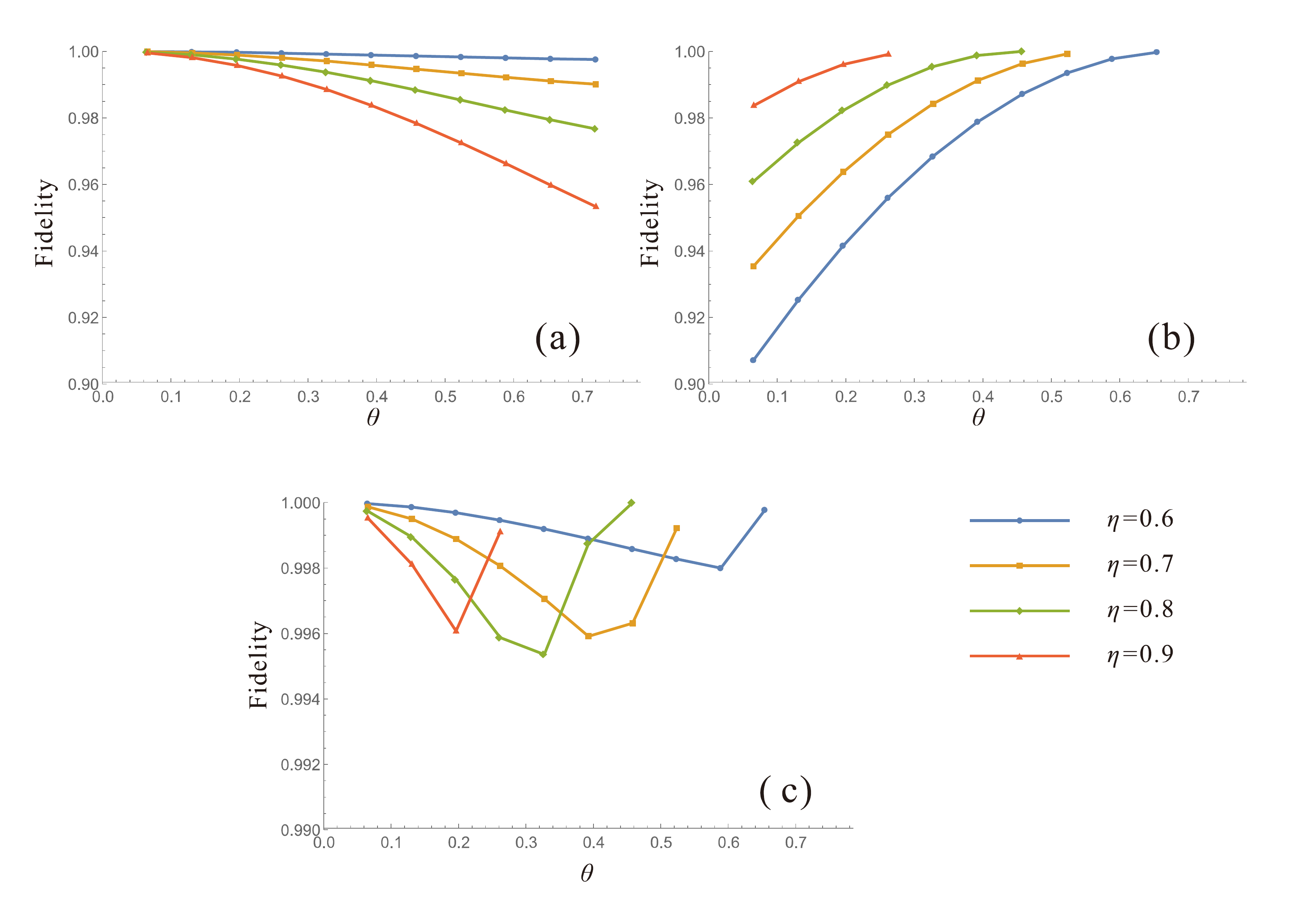}\\
\caption{The minimum fidelities to determine the corresponding assemblages to be one-way steerable. (a) The minimum fidelities of Bob's CSs to make sure that Alice can steer Bob's state. (b) The minimum fidelities of Alice's CSs to ensure that Alice's state cannot be steered by Bob. (c) The minimum fidelities of Alice's and Bob's CSs to make sure one-way steering.}\label{figr4}
\end{figure}

Now let us consider the fidelity of the CSs on each side. As shown in Fig. \ref{figr3}, for a given state $\rho_{theo}$ and a fidelity value, the possible locations of CSs are restricted in the corresponding red regions. In the case of the steering process from Alice to Bob, one need to prove that for any Bob's possible assemblage, there isn't any LHSM description as shown in Fig. \ref{figr3} (a). By numerical calculation, the minimum fidelity to determine the assemblage to be steerable is obtained which is denoted as $F_{Bob}$. Considering some entangled states, the corresponding minimum fidelities are shown in Fig. \ref{figr4} (a). If the minimum fidelity of Bob's four experimental CSs is higher than $F_{Bob}$, Bob has to conclude that Alice can steer his state. Conversely, to ensure that Bob could not steer Alice's state, all of Alice's possible assemblages should have the LHSM descriptions as shown in Fig. \ref{figr3} (b). By numerical calculation, the minimum fidelity ($F_{Alice}$) to ensure the assemblage to be not steerable is obtained as shown in Fig. \ref{figr4} (b). If the minimum fidelity of Alice's four experimental CSs is higher than $F_{Alice}$ , Bob fails to steer Alice's state. To ensure the one-way steering, the minimum fidelities of Alice's and Bob's CSs should be larger than $\max\{F_{Alice},\,F_{Bob}\}$ as shown in Fig. \ref{figr4} (c).

\begin{figure}[h]
\includegraphics[width=0.45\textwidth]{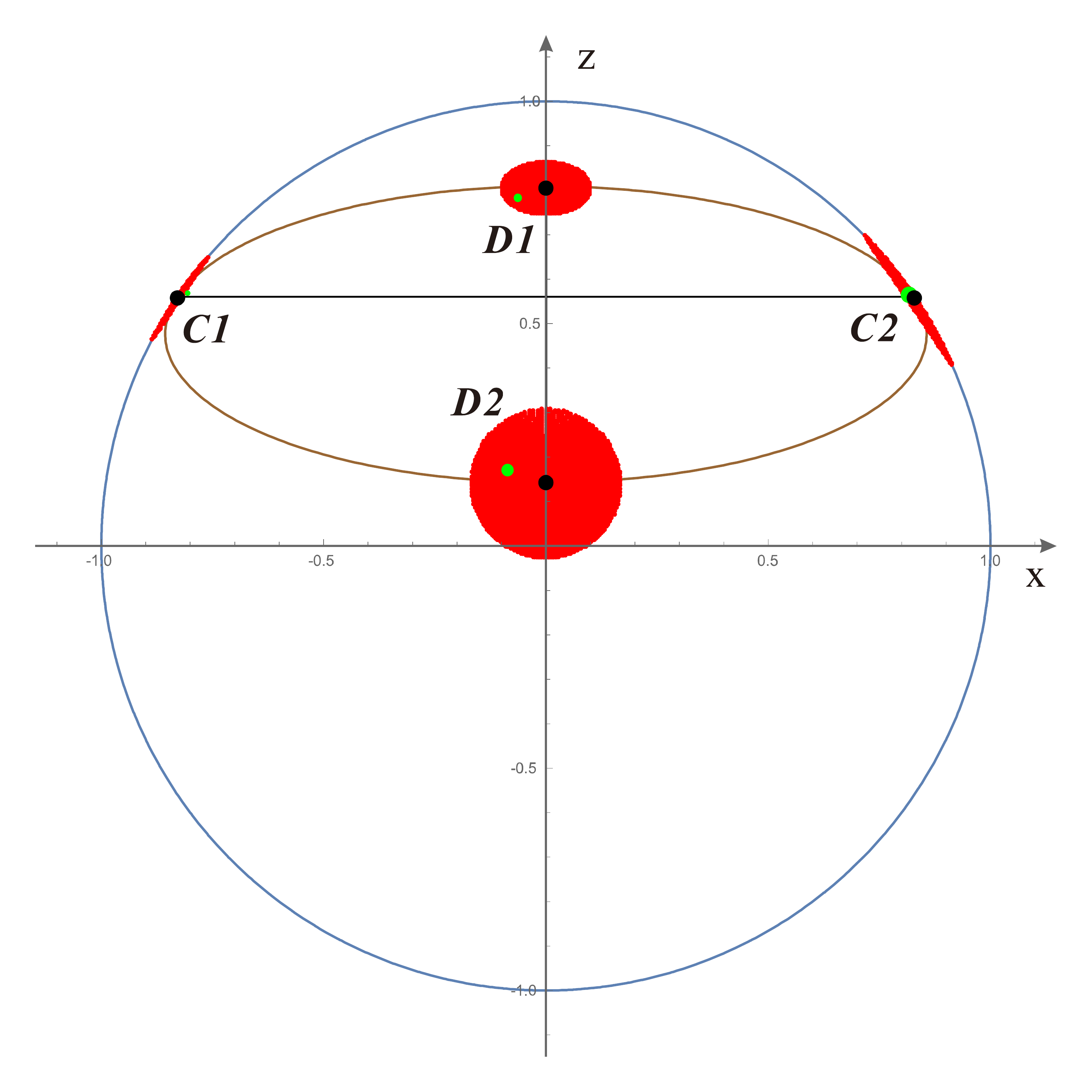}\\
\caption{Illustration for using the fidelity to calculate steering radius for the entangled state with $\theta=0.488$ and $\eta=0.727$. The brown ellipse represents Bob's CSs when Alice measures along the direction in $XZ$ plane. The black points represent Bob's theoretical CSs when Alice measures along $x$ and $z$. The four red regions represent the states whose fidelities are larger than the corresponding experimental results. The fidelities of the states in regions C1, C2, D1, D2 are larger than 0.997, 0.992, 0.998, 0.993, respectively. The green circles represent the possible experimental CSs based on the coordinate errors.}\label{figr5}
\end{figure}

In our work, the calculation of steering radius only depends on Bob's own measurement results in the case of the steering process from Alice to Bob. The experimental errors are considered to be the coordinate errors of CSs in the Bloch sphere, which are deduced from the counting statistics. Here, We further consider the case using the experimental fidelities of the CSs to calculate the steering radius. One of the examples is shown in Fig. \ref{figr5}. The four red regions represent the states whose fidelities, compared with the theoretical states shown by the black points, are larger than the corresponding experimental results. Due to the fact that the regions of the possible experimental CSs are magnified when using the fidelity, the corresponding steering radius is calculated to be smaller than the one we obtained in this work. It is likely that the steering radius of some steerable states are calculated to be below unity when using the fidelity, i.e., the steering process fails.

\section{Preparation of the local hidden states}
For any normalized single-qubit state, its coordinate in the Bloch sphere could be written as $G=(G_x,\,G_y,\,G_z)$. Thus the state's matrix can be written as
\begin{equation}\label{method1}
\rho_G=\frac{I+G\cdot\vec{\sigma}}{2}=\frac{1}{2}\left(
\begin{array}{cc}
1+G_z & G_x-i G_y\\
G_x+i G_y & 1-G_z\\
\end{array}
\right).
\end{equation}
As mentioned in the text, the local hidden states are always located in the $XZ$ plane of the Bloch sphere when the two measurement settings are $x$ and $z$, i.e., $G_y=0$. Thus, we can rewrite the state's matrix as $\rho_G=\mu|\phi_\alpha\rangle\langle \phi_\alpha|+\nu|H\rangle\langle H|$ or $\rho_G=\mu|\phi_\alpha\rangle\langle \phi_\alpha|+\nu|V\rangle\langle V|$, where $|\phi_\alpha\rangle=\cos\alpha|H\rangle+\sin\alpha|V\rangle$ and $\mu^2+\nu^2=1$. We can then obtain the unknown parameters $\alpha$, $\mu$ and $\nu$. By rotating the half-wave plates (HWP) and adjusting the removable shutters in the unbalanced Mach-Zehnder interferometers (UMZ) shown in Fig. 2. {\bf (b)} and {\bf (c)}, we can prepare the desired local hidden states. In the experiment, $G_y\neq0$ and the hidden states can be easily prepared with a quarter-wave plate following the HWP in UMZ.

\subsection{More experimental results}
 In our experiment, three states are prepared to realize the two-way steering task, and the values are shown in Table \ref{table}.

\begin{table}[!ht]
\renewcommand\arraystretch{1.5}
\caption{The detailed values of $R$ for the three states realizing two-side EPR steering process.}\label{table}
\centering
\begin{tabular}{|p{2cm}<{\centering}|p{2cm}<{\centering}|p{1.5cm}<{\centering}|p{1.5cm}<{\centering}|}\hline
$\theta$ & $\eta$   & $R(\rho_{AB})$ & $R(\rho_{BA})$ \\ \hline
0.775(5) & 0.989(1) & 1.36     & 1.35 \\ \hline
0.773(6) & 0.796(4) & 1.13     & 1.09 \\ \hline
0.536(4) & 0.816(4) & 1.11     & 1.07 \\ \hline
\end{tabular}
\end{table}

\begin{figure}[h]
\includegraphics[width=0.45\textwidth]{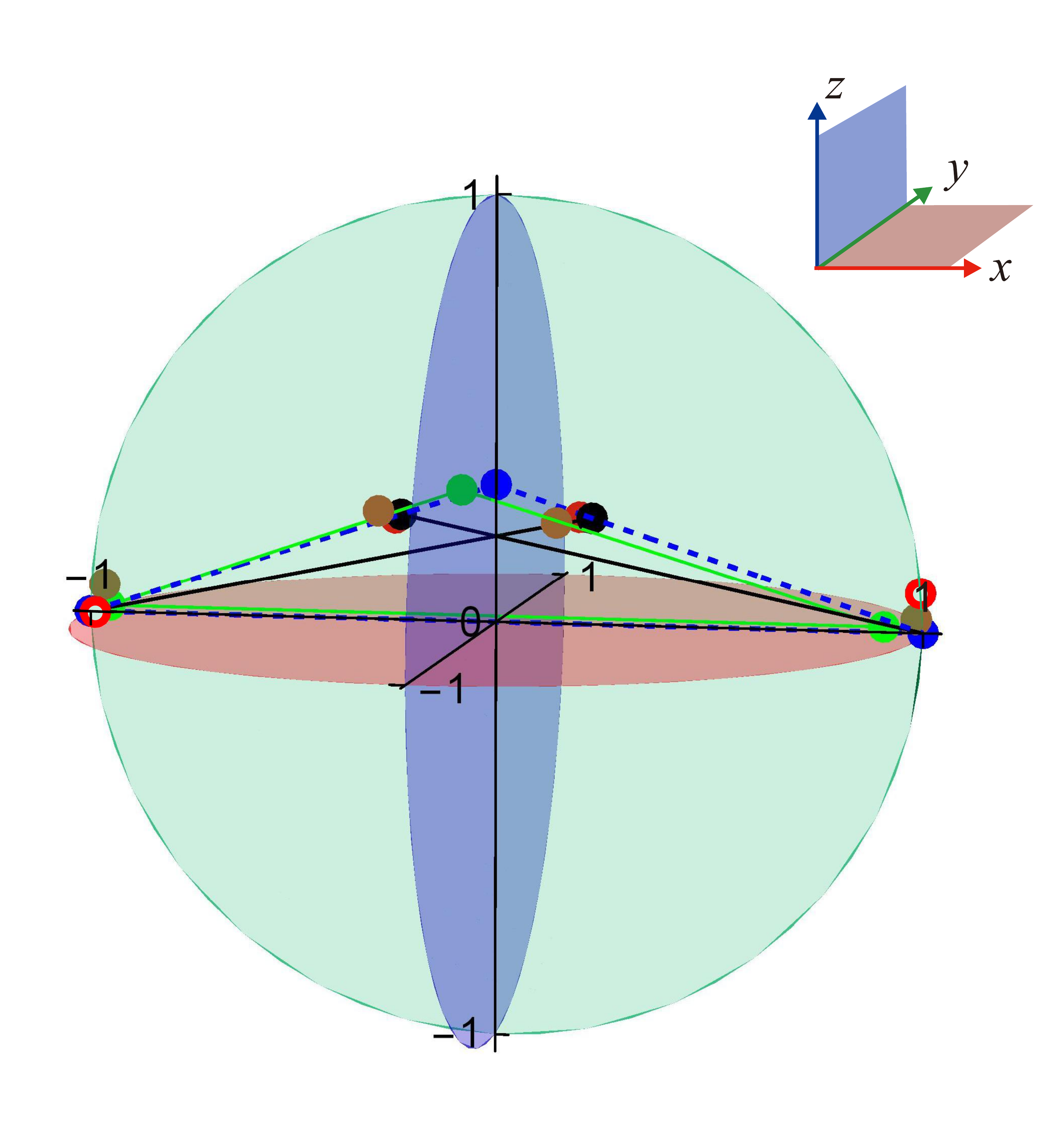}\\
\caption{Experimental results for three local hidden states reconstructing Alice's four CSs for the state with $\theta=0.328$ and $\eta=0.629$. The three blue and three green points are the hidden states in theory and experiment, respectively. The black, red hollow and brown points represent the theoretical results, experimental results and the simulated results with hidden states of the normalized CSs, respectively.}\label{figs2}
\end{figure}

When Bob steers Alice's state along two measurement settings, there is a singular situation where only three hidden states are needed to simulate Alice's four conditional states (CSs). In this case, one of Alice's two normalized CSs is a pure state when Bob measures along the two directions $\vec{n}_1 = (\sin2\theta, 0, \cos2\theta)$, $\vec{n}_2 = (-\sin2\theta, 0, \cos2\theta)$. The pure state $1/\sqrt{2}|H - V\rangle$ ($1/\sqrt{2}|H + V\rangle$) is one of Alice's two CSs when the direction is $\vec{n}_1$ ($\vec{n}_2$) performed by Bob. The other state is a mixed state. Fig. \ref{figs2} shows the corresponding experimental results, where the Alice's two CSs are linked by the black line.

\begin{figure}[h]
\includegraphics[width=0.45\textwidth]{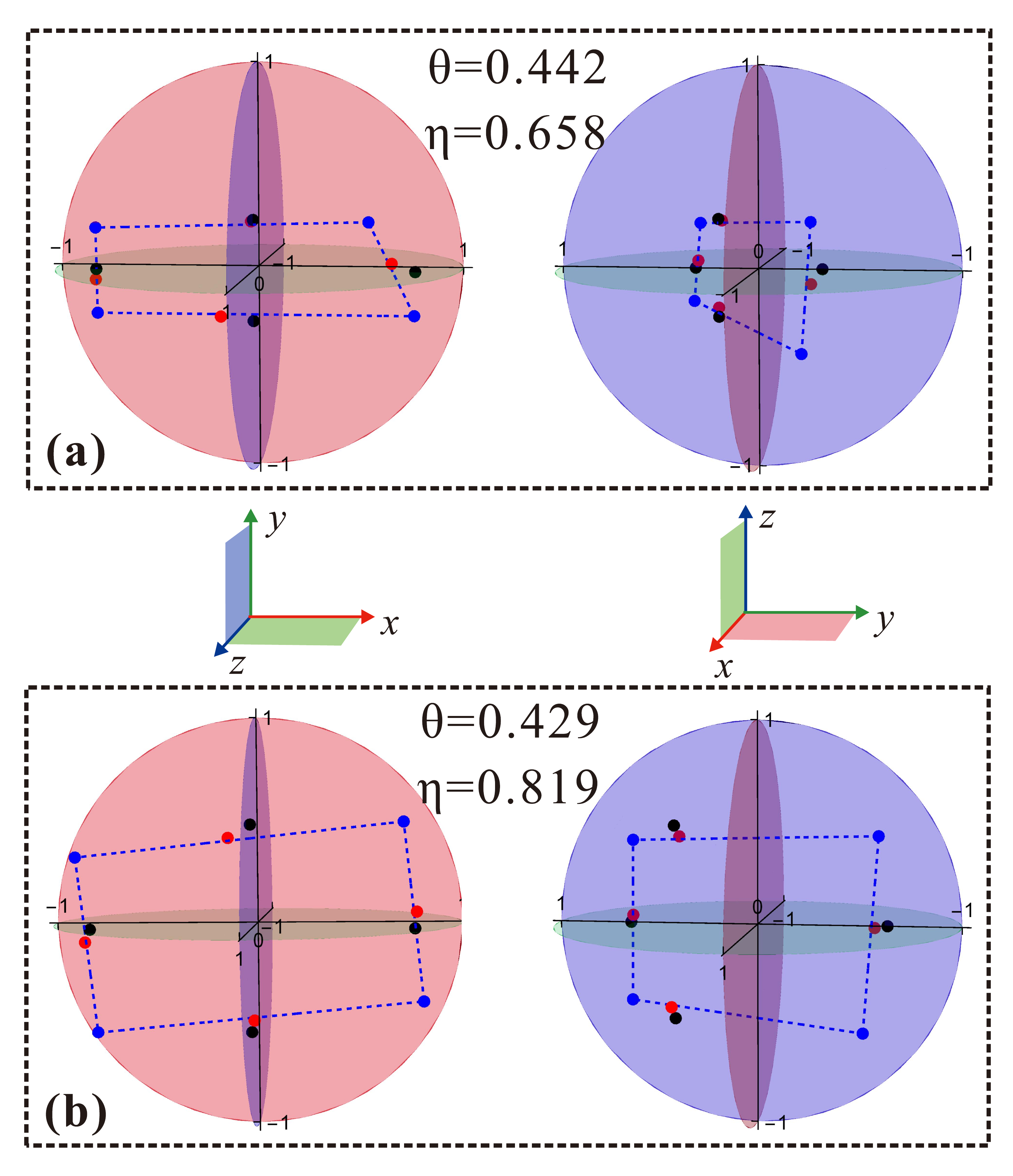}\\
\caption{Experimental results of Alice's states in the Bloch sphere when Bob measures along  $\{x,y\}$ and $\{y,z\}$. The states they shared are {\bf (a)} $\{\theta=0.442\pi, \eta=0.658\}$ and {\bf (b)} $\{\theta=0.429,\eta=0.819\}$. The theoretical and experimental results of normalized CSs are marked by the black and red points, respectively. The blue points represent the theoretical results of the four local hidden states.}\label{figs3}
\end{figure}

According to the discussion above, the steering task from Bob to Alice fails when $|\cos2\theta|\geq|2\eta-1|$. In the experiment, Bob measures along $x$ and $z$. If Alice finds an LHSM, then $|\cos2\theta|\geq|2\eta-1|$, which is confirmed experimentally. In the case of $|\cos2\theta|\geq|2\eta-1|$, Bob further measures along the sets of $\{x,\,y\}$ and $\{y,\,z\}$ to confirm that there still exist LHSMs for Alice. The experimental results are shown in Fig. \ref{figs3}.

\end{document}